\documentclass[conference]{IEEEtran}
\IEEEoverridecommandlockouts
\pdfoutput=1
\usepackage{cite}
\usepackage{amsmath,amssymb,amsfonts}
\usepackage{algorithmic}
\usepackage{graphicx}
\usepackage{textcomp}
\usepackage{threeparttable}
\usepackage{xcolor}
\usepackage{url}
\usepackage{multirow}
\usepackage{hyperref}
\usepackage{listings}

\lstdefinestyle{mystyle}{
    language=Python,
    basicstyle=\ttfamily\footnotesize,    
    keywordstyle=\color{blue}\bfseries,   
    commentstyle=\color{gray},            
    stringstyle=\color{red},              
    numberstyle=\tiny\color{teal},        
    stepnumber=1,                          
    showstringspaces=false,                
    breaklines=true,                        
    frame=single,                           
    backgroundcolor=\color{white},         
    morekeywords={self, True, False, None}  
}
\definecolor{myblue}{RGB}{0, 51, 102} 

\usepackage{booktabs, multirow}
\usepackage{pifont}
\usepackage[breakable, skins]{tcolorbox}

\usepackage[breakable, skins]{tcolorbox}
\tcbuselibrary{skins}
\newenvironment{summarybox}
{\begin{tcolorbox}
[breakable,enhanced,arc=0mm,colback=gray!10,frame hidden,overlay broken={%
    \draw[thick,black] (interior.north west)--(interior.south west);
},overlay unbroken={%
    \draw[thick,black] (interior.north west)--(interior.south west);
},left=2pt,right=0pt,top=0pt,bottom=0pt,before={\vspace{3pt}\noindent},after={\vspace{0pt}}]
\setlength{\baselineskip}{0.75\baselineskip}}
{\end{tcolorbox}}
\newenvironment{summary}
{\vspace{5pt}\noindent\begin{summarybox}}
{\end{summarybox}\vspace{-5pt}}

\begin{document}

\title{AI-Powered, But Power-Hungry? Energy Efficiency of LLM-Generated Code
}

\author{\IEEEauthorblockN{Lola Solovyeva}
\IEEEauthorblockA{\textit{University of Twente} \\
Enschede, The Netherlands \\
o.solovyeva@utwente.nl}
\and
\IEEEauthorblockN{Sophie Weidmann}
\IEEEauthorblockA{\textit{University of Twente} \\
Enschede, The Netherlands \\
s.weidmann@student.utwente.nl}
\and
\IEEEauthorblockN{Fernando Castor}
\IEEEauthorblockA{
\textit{University of Twente}\\
Enschede, The Netherlands \\
f.castor@utwente.nl}}

\maketitle

\begin{abstract}
Large language models (LLMs) are used in software development to assist in various tasks, e.g., code generation and code completion, but empirical evaluations of the quality of the results produced by these models focus on correctness and ignore other relevant aspects, such as their performance and energy efficiency.  Studying the performance of LLM-produced programs is essential to understand how well LLMs can support the construction of performance- and energy-critical software, such as operating systems, servers, and mobile applications. This paper presents the first study analyzing the energy efficiency and performance of LLM-generated code for three programming languages Python, Java, and C++, on two platforms, a Mac and a PC, leveraging three frontier LLMs, Github Copilot, GPT-4o, and the recently-released OpenAI o1-mini, and targeting ``hard'' programming problems from LeetCode. Our results show that the models are much more successful in generating Python and Java than C++ code.
Also, LLM-generated code sometimes surpasses an efficient human-written solution, although that is language-dependent and the language with the best results, Python, is the one where application performance and energy consumption tend to matter the least in practice. Furthermore, the performance of generated code is highly correlated across the two platforms, hinting at potential for results to be portable across platforms.

\end{abstract}


\section{Introduction}
Among rapid technological advancements, growing interest in the environmental impact of software development has led to thorough investigations into its energy consumption, resource utilization and carbon emissions~\cite{KERN201553, CAPRA201260, EmpericalGreen}. Software development processes frequently involve the extensive use of energy-extensive resources, such as servers, storage systems, and networking infrastructure. The aforementioned resources contribute to substantial carbon emissions and resource depletion~\cite{10214579}. Data centers are responsible for approximately 1\% of global energy consumption and contribute an estimated 2–4\% of worldwide carbon emissions~\cite{ZHU2023104322}. Considering the energy demands of the ICT sector, it accounted for approximately 4\% of global electricity consumption during its operational phase and contributed around 1.4\% of global greenhouse gas emissions in 2020~\cite{MALMODIN2024102701}.

Current recent advancements in the aforementioned areas such as enhanced cooling systems and the transition from local processors to large-scale data centers have contributed to reducing energy demands. However, advances in machine learning (ML) and artificial intelligence (AI), where demand remains consistently high, may exceed these efficiency gains~\cite{vartziotis2024learn, ALZOUBI2024143090, 8584429}. ML has already found application in numerous areas and continues demonstrating considerable potential across multiple industry sectors and aspects of life, including healthcare, agriculture, engineering, finance, gaming, and transportation. The lifecycle emissions of an AI model include emissions produced during training and testing phases, as well as those arising from the inference during its deployment~\cite{DESISLAVOV2023100857}. To put it in perspective, 
just training BLOOM, a 176B parameter language model, required an estimated 689,842 KWh~\cite{JMLR:v24:23-0069}, approximately the energy consumed by 1000 Tesla Model 3 cars running for almost 5,000 km each. Its training emitted an estimated 24.7 tonnes of CO2~\cite{JMLR:v24:23-0069}, the emissions of a 737 flying between Rome and London with 100 passengers.

\par Although the development and deployment of AI models are associated with a substantial carbon footprint~\cite{10.1007/978-3-031-73110-5_22}, they can have a potential to promote environmental sustainability~\cite{BOLONCANEDO2024128096}. For example, one of the applications of generative AI in software development is code generation. The efficiency of the code, including factors such as energy consumption and carbon footprint, remains crucial, despite being neglected by the programmers themselves during the development process. With the integration of the AI-assisted tools, such as GitHub Copilot, which aim to facilitate ``faster and smarter''\footnote{\url{https://code.visualstudio.com/docs/copilot/overview}} code development, it becomes necessary for them to also account for the efficiency of the solutions they generate or suggest. By generating energy-efficient code, LLMs have potential to reduce the carbon footprint and enable resource saving of the produced software, particularly  for compute-intensive tasks, where the efficiency of the generated code may outweigh the costs of generating it. 
\par This study seeks to assess the energy efficiency of code generated by three LLMs, GPT-4o, OpenAI o1-mini, and Github Copilot, providing insights into how closely LLM-generated solutions align with the efficiency of human-written code. It examines coding tasks including  non-trivial techniques, such as greedy algorithmic techniques, graph algorithms, and numerical computation, among others. The programming problems are sourced from LeetCode, which is a platform that is widely utilized for assessing the programming capabilities of LLMs. Furthermore, the chosen problems are specifically classified as ``hard.'', which has two implications: (i) the analyzed LLMs were not directly trained on them, and (ii) they are generally considered challenging for humans. Also, it is the first study to determine whether the results hold consistently across multiple programming languages and two machines operating on distinct systems, Ubuntu and macOS Sonoma.
\par Our findings reveal that LLMs perform optimally in Python, achieving the highest pass@1 accuracy. In terms of energy efficiency, the models demonstrate results comparable to the baseline for Python and Java, with Python solutions, in some instances, exhibiting greater energy efficiency. Tasks related to \textit{String}, \textit{Tree}, \textit{Hashing}, and \textit{Search algorithms} consistently show strong performance, while challenges persist in \textit{Sorting}, \textit{Graph}, \textit{Greedy algorithms}, and tasks involving \textit{Math} and \textit{Recursion}, resulting in more energy-demanding solutions. The OpenAI o1-mini model shows significant improvements in accuracy, particularly in \textit{Search algorithms} and \textit{Sorting}, but it exhibits higher energy consumption compared to the earlier models, GPT-4o and GitHub Copilot. Lastly, our findings showed that LLM-generated solutions are machine-agnostic with strong energy correlations across systems.

The replication package for this study, along with the appendices, is publicly available~\cite{EnergyEfficiencyLLMCode2024}.
\section{Related Work}
\label{sec:relatedWork}
Previously, most studies solely focused on evaluating correctness of the code generation\cite{10403378, 10.1145/3627217.3627233, Chen2021EvaluatingLL}. However, the focus has shifted to also addressing the issue of efficiency in code generated by LLMs as AI-assisted tools become increasingly prevalent in software engineering and development processes~\cite{han-etal-2023-sample, Nitin_Sherje_2024, copilotAcc}.
\\
A closely related study by Vartziotis et al.~\cite{vartziotis2024learn} examines the energy efficiency of Python code generated by three widely used tools, namely GitHub Copilot, ChatGPT 3 and Amazon CodeWhisperer. The results show that AI models can generate code optimized for sustainability when explicitly requested to do so. However, they also reported that  human-written code is consistantly more energy-efficient.  A related study by Coignion et al.~\cite{10.1145/3661167.3661221} evaluates LLM-generated code from a performance perspective. Their analysis compares 18 LLMs using LeetCode data, examining factors such as model temperature and success rate and their influence on code performance. The findings of this study align with those of Varziotis et al.~\cite{vartziotis2024learn}, emphasizing that LLM-generated code, on average, demonstrates greater efficiency compared to human-written code. Both studies, however, share limitations, as they focus exclusively on Python data, with Varziotis et al.~\cite{vartziotis2024learn} deriving their conclusions from a limited dataset comprising only six coding problems.
\par In contrast, this study investigates three programming languages: Python, Java, and C++. Furthermore,it utilizes a comprehensive benchmark of 53 coding tasks, thereby increasing the robustness and reliability of the findings. We argue that Python is generally not regarded as an inherently efficient programming language~\cite{PEREIRA2021102609}, making it less likely to be chosen by developers when performance is a critical requirement. Consequently, studies evaluating the performance and energy efficiency of LLM-generated code for Python may have limited practical relevance, as such analyses may not align with real-world scenarios where more performance-oriented languages are typically preferred.
\par Several studies~\cite{du2024mercury, huang2024effibench, 10.1145/3597503.3623316, Hendrycks2021MeasuringCC, wang-etal-2023-recode, Huang2024EffiCodeUC} have introduced benchmarks specifically designed to evaluate LLM-generated code, focusing on runtime performance and memory consumption. In contrast, our study expands this scope by including the energy consumption as an additional metric, providing a more comprehensive overview of the energy efficiency of LLM-generated code. Furthermore, while the benchmarks developed by two of the prior studies~\cite{du2024mercury, huang2024effibench} were extensive, each encompassing over 1000 coding problems, their analyses were limited to Python, whereas our study also includes Java and C++. 
\par Rather than developing a benchmark from scratch, the study by Liu et al.~\cite{liu2024evaluatinglanguagemodelsefficient} grouped efficiency-demanding Python programming tasks from \textit{HumanEval+} and \textit{MBPP+} to form \textit{EvalPerf} addressing the limitation of previous works, which primarily focused on light computational requirements and possibly misrepresenting the capabilities of LLMs. To improve the quality of their evaluation, they augmented their experiments with computationally intensive inputs, aiming to provide a more accurate assessment of the efficiency and performance of the generated code. Although the referenced study focused exclusively on performance-intensive tasks from \textit{HumanEval+} and \textit{MBPP+}, several other works have shown that LLMs perform notably well on these datasets. In contrast, our research does not depend on these benchmarks, as they are generally labeled as "Easy" and are less challenging for evaluating model capabilities.  In addition, we aim to compare the efficiency of the generated code with human-written solutions, a consideration that was overlooked in the cited study.
\par The study by Du et al.~\cite{duCodeGen} sought to examine a more complex code generation scenario. To this end, they developed their own benchmark, \textit{ClassEval}, which consists of 100 class-level Python code generation tasks. Based on the new benchmark, this is the first study that evaluated LLMs in the context of class-level code generation. Their experiments included 11 state-of-the-art models, each varying in size, architecture, data sources, and application domains. The objective of the cited study differs from the primary goal of our research. While the cited work introduced a novel benchmark to assess the correctness of the code generated by LLMs, our study focuses primarily on evaluating the energy efficiency of the generated code. Nonetheless, we also account for the correctness of the code, as evaluating the efficiency is only meaningful when the code is correct.
\par LeetCode, although primarily a platform for coding competitions,  is also extensively used as a dataset for evaluating the programming capabilities of LLMs. Döderlein et al.~\cite{Dderlein2022PilotingCA} evaluated the performance of Copilot and Codex on LeetCode, analyzing the impact of varying prompt structures on the models' effectiveness. Nguyen and Nadi~\cite{9796235} investigated GitHub Copilot's code recommendations for LeetCode problems, focusing on the complexity and intricacies of the generated solutions. Vasconcelos et al.~\cite{10.1145/3702320} examined the impact of emphasizing uncertainty in AI-driven code completions, utilizing LeetCode problems and the Codex model as part of their analysis.

\section{Methodology}
\label{sec:methodology}
By utilizing the formulation proposed by Basili et al.~\cite{basil}, the high-level goal of this study is to \textit{analyze} LLM-generated code \textit{for the purpose of} evaluation \textit{with respect to their} energy-efficiency \textit{from the viewpoint of} software developers \textit{in the context of} Python, C++ and Java based applications. 
\par Our high-level goal can be summarized in the following primary research question: 
    \begin{quote}
        \emph{RQ: To what extent can energy-efficient code be achieved via utilization of Large Language Models (LLMs)?}
    \end{quote}
To address the primary research question, we evaluate and compare the energy efficiency of code generated by LLMs against human-written solutions that are considered efficient. To gain a more comprehensive understanding and conduct an in-depth examination of the topic, the primary research question is divided into the following sub questions:\\
\textbf{RQ1: }\textit{What are the variations in energy-efficiency of the LLM-generated code across different programming languages?} Python remains the primary language for evaluating the capabilities of code generated by LLMs. However, other prominent and widely-used programming languages have yet to be thoroughly explored. We hypothesize that the energy efficiency of LLM-generated code may vary across different programming languages, when compared to human-written solutions, potentially due to the diversity of samples present in the training data for each language, and the languages' particularities. The goal here is to compare LLM-generated and human-written solutions across three programming languages. Our goal is not to compare programs in different programming languages directly. 
\\
\textbf{RQ2: }\textit{What is the impact of data structure and algorithmic technique selection on the energy efficiency of LLM-generated code?} Software development encompasses a wide range of programming algorithms to tackle various tasks, including recursion, search and sorting strategies, among others. The complexity of these algorithms may vary, with some being easier to optimize than others. Accordingly, this research question focuses on identifying some characteristics of algorithmic techniques and data structures that pose greater challenges for LLMs in generating energy-efficient solutions, while also exploring potential reasons for these difficulties. \\
\textbf{RQ3: }\textit{Is the energy efficiency of programs generated by different LLMs significantly different?} Advancements in developing LLMs, that could handle increasingly complex tasks, have maintained a strong emphasis on improving the correctness of code generation. However, this sub-research question seeks to investigate whether energy efficiency can also be taken into account as a distinguishing factor among different models. \\
\textbf{RQ4: }\textit{Is there a significant difference in the energy efficiency of LLM-generated programs across different platforms?
}
The energy efficiency of LLM-generated code is not solely determined by the code itself but is also influenced by the platform on which it runs, including the operating system and underlying hardware. As such, the choice of platform can significantly affect the energy footprint of the same program. This question seeks to examine whether LLM-generated code demonstrates variations in efficiency when executed on a specific system and whether improvements (or deterioration) in energy efficiency promoted by LLMs have the potential to be transferable across platforms.

\subsection{Baseline}
Our benchmark for the baseline is built from the human-written solutions of ``Hard'' programming tasks posted on LeetCode. Many studies, including those of Vartziotis et al.~\cite{vartziotis2024learn} and Niu et al.~\cite{niu2024evaluating}, have used LeetCode as a benchmark, since it provides a wide range of coding problems and uses a community voting system to rank solutions, making it a reliable source for efficient human-written code. In LeetCode, a "Hard" problem designation indicates that the problem poses stringent constraints on time and space complexity, necessitating both advanced intuition and a thorough understanding of data structures. Each programming problem includes a tag indicating the specific method or data structure employed in its construction. This tag is selected by the LeetCode maintainers as part of their curation process. A tag is subsequently used to categorize the solution in terms of the relevant data structures and algorithmic techniques, e.g., greedy, recursion, dynamic programming, etc. Additionally, a single problem may have multiple tags, allowing it to be classified under multiple groups. The list of tags can be seen in Table \ref{tab:tags_total_accuracy}. It is important to note that certain tags were combined into a single category, as they either perform similar algorithmic techniques or are associated with the same data structure. For instance, Depth-First Search and Breadth-First Search were both classified under the \textit{Search algorithms} group.
\par Each problem includes three human-written solutions, one for each programming language: Python, Java, and C++. They were chosen based on the upvotes provided by the users of the platform. Additionally, the authors outlined the time and space complexity of their solutions, aiming to minimize these metrics in accordance with LeetCode's acceptance criteria. Hence, the solutions in the benchmark represent very efficient solutions to a given problem. Thus, for 53 problems, our benchmark comprises 159 solutions, which serve as a baseline for comparison against solutions generated by LLMs. 

\subsection{Variables}
This study involves the following independent and dependent variables:
\begin{itemize}
    \item \textbf{Independent Variables}:

    \textbf{Source of the code (LLM-generated vs. human-written)}: 
        This variable distinguishes between whether the code was generated by a Large Language Model or written by a human. Three LLMs are evaluated in this study, namely GitHub Copilot, ChatGPT 4o, and OpenAI o1-mini. 
        
    \textbf{Programming language (Python, Java, C++)}: 
        Three programming languages are observed, namely Python, Java, and C++, which have different performance and efficiency characteristics. Python is slower but flexible, Java is balanced, and C++ is known for its efficiency. 
        

        \textbf{System type}: 
        Different hardware systems or operating environments can influence how efficiently the code runs. This variable accounts for the differences in system configurations (e.g., CPU architecture, available memory, power management settings) that may impact energy consumption, execution time, and resource usage.\\

    \item \textbf{Dependent Variables}:

         \textbf{Energy consumption (in Joules)}: 
        This measures the total energy consumed during code execution, focusing on CPU energy.  Lower energy consumption implies greater efficiency and reduced environmental impact. 
        
        \textbf{Execution runtime (in milliseconds)}: 
        The time taken to complete a task, measured in milliseconds. Lower execution times indicate better performance.

        \textbf{Correctness of generation (in \%)}:
        The number of generated solutions that passed pre-defined test sets. The value is expressed as the number of generated solutions that passed the tests over total number of generated solutions.  
\end{itemize}
\subsection{Workloads, Prompts and Benchmark}
Our benchmark for evaluating LLM-generated code encompasses solutions for 53 problems implemented in three programming languages, produced by three different models. Consequently, the benchmark includes a total of 477 solutions. Each model was provided with the problem description and the corresponding type signature, both taken from LeetCode, as input prompts. 
\par The workloads in this study refer to difficult programming tasks obtained from LeetCode to test the performance and efficiency of both LLM-generated and human-written code. Their tests serve as the basis for measuring performance metrics in regards to energy efficiency. The input for each test is taken from the list of examples provided by LeetCode for the specific problem in question. Even though they might be considered as light computation, we believe they effectively simulate the average real-world usage of these problems. Additionally, supplementary test cases were incorporated to evaluate edge scenarios involving maximal possible input, thereby increasing the computational workload on the programs.   \\ 
The types of programming tasks in the workloads could be found in Table \ref{tab:tags_total_accuracy},which inherently vary in computational complexity and resource demands. Each code solution is executed under identical workload conditions to ensure fair comparisons. The same set of tests, written in the respective programming language, is then run on both LLM-generated and human-written code. These tests are executed sequentially 10 times to account for performance variations and to collect reliable data.
\subsection{Design}
This study aims to compare the energy-efficiency of solutions generated by LLMs with human-written solutions that are optimized based on space and time complexities. 
To perform a comparison, we generate solutions using GitHub Copilot, ChatGPT 4o, and OpenAI o1-mini for three programming languages: Python, Java, and C++. Hence, we have 9 programs for the measurements. The selection of these languages is justified based on their widespread usage and distinctive roles: Python, as a popular scripting language, is extensively utilized in data science and machine learning; Java, as a versatile managed language, is employed across various domains, from mobile applications to server-side development; and C++, as a systems programming language, is favored for its emphasis on high performance. The models are prompted with instructions detailing the task requirements and the relevant type signatures. 
The generated solutions are then evaluated for correctness using predefined test sets specific to each task. The test set includes cases from LeetCode and additional tests for edge cases, such as maximum and minimal input, which we created and validated against the baseline code to ensure their correctness and comprehensive coverage. The average number of test cases per programming problem is 7. We discard the LLM-generated solutions that do not pass the tests. Then, dependent variables are recorded across two systems, as mentioned in Table \ref{tab:hardware}, for both LLM-generated solutions and the baseline. This selection of platforms is based on their widespread popularity, extensive usage, and the availability of reliable tools for accurate measurements. As each of the nine programs is executed on two systems, a total of 18 measurements are recorded for a single benchmark. Finally, the results are compared and evaluated using pass@1 accuracy, energy consumption and time of code execution. 

\subsection{Measurement Environment}
To ensure accurate and reproducible results, the environment in which the experiment is conducted is defined precisely, taking into account both hardware and software elements that may affect the results. Experiments were conducted on two platforms with different configurations to capture the energy efficiency of different hardware conditions. The hardware specifications of each system can be found in the Table \ref{tab:hardware}. For simplicity, we refer to the Apple MacBook Air and the Lenovo Thinkpad simply as the macOS and Ubuntu systems (or machines), respectively.
\begin{table}[!t]
\centering
\caption{Specifications of the hardware used in the experiments.}
\begin{tabular}{|l|c|c|}
\hline
\textbf{Name}          & Apple MacBook Air                                                          & \begin{tabular}[c]{@{}c@{}}Lenovo ThinkPad \\ P16v Gen2\end{tabular}           \\ \hline
\textbf{Processor}     & M3 chip                                                                    & \begin{tabular}[c]{@{}c@{}}13th Gen \\ Intel Core i7\end{tabular}              \\ \hline
\textbf{Cores}         & \begin{tabular}[c]{@{}c@{}}4 P-core \\ 4 E-core\end{tabular}               & \begin{tabular}[c]{@{}c@{}}8 P-core \\ 12 E-core\end{tabular}                  \\ \hline
\textbf{Max Frequency} & \begin{tabular}[c]{@{}c@{}}P-core 4.06 GHz \\  E-core 2.75GHz\end{tabular} & \begin{tabular}[c]{@{}c@{}}P-core 5.3GHz \\  E-core 3.8GHz\end{tabular}        \\ \hline
\textbf{RAM}           & 16GB                                                                       & 32GB                                                                           \\ \hline
\textbf{SSD}           & 256GB                                                                      & 1T                                                                             \\ \hline
\textbf{GPU}           & M3 10-core                                                                 & \multicolumn{1}{l|}{NVIDIA RTX 3500}                                           \\ \hline
\textbf{OS}            & macOS Sonoma v14.6                                                         & \begin{tabular}[c]{@{}c@{}}Ubuntu 22.04.5 LTS\\ kernel v 6.8.0-45\end{tabular} \\ \hline
\end{tabular}
\label{tab:hardware}
\end{table}
\par The code was compiled using the appropriate compilers for each programming language: GCC 11.4.0 for C++, OpenJDK23 for Java, and Python 3.13.0 interpreter. The choice of compilers and interpreters is based on their compatibility with the human-written solutions, as some of these solutions were submitted several years ago and may not be compatible with recent updates in software packages. Default compiler optimization flag -O0 is set for all of the executions. All executions are conducted via the terminal window on both systems under identical conditions, including being plugged in to the power outlet and fully charged, disabled Wi-Fi and Bluetooth, and no other applications or browsers running in the background. Furthermore, Linux runs happened under the default ondemand governor\footnote{\url{https://www.kernel.org/doc/Documentation/cpu-freq/governors.txt}}.
\begin{table*}[!t]
\centering
\caption{Summary of results for all programming languages, models, and machines analyzed in this study for pass@1 accuracy, energy consumed by the machine, and execution time. Energy consumption and execution time are reported as percentages relative to the baseline, with raw measurements taken in joules and seconds, respectively. The model names have been abbreviated in the table for clarity, with OpenAI o1-mini referred to as \textbf{o1}, GPT-4o as \textbf{4o}, and GitHub Copilot as \textbf{copilot}.  }
\label{tab:accuracy_energy_exec}
\scriptsize
\begin{tabular}{|l|ccc|cccccc|cccccc|}
\hline
\multirow{3}{*}{}                     & \multicolumn{3}{c|}{\multirow{2}{*}{\textbf{Accuracy}$\uparrow$}}                                & \multicolumn{6}{c|}{\textbf{Ubuntu}}                                                                                                                                                                 & \multicolumn{6}{c|}{\textbf{macOS}}                                                                                                                                                                  \\ \cline{5-16} 
                                      & \multicolumn{3}{c|}{}                                                                  & \multicolumn{3}{c|}{\textbf{Energy}$\downarrow$}                                                                        & \multicolumn{3}{c|}{\textbf{Execution Time}$\downarrow$}                                           & \multicolumn{3}{c|}{\textbf{Energy}$\downarrow$}                                                                        & \multicolumn{3}{c|}{\textbf{Execution Time}$\downarrow$}                                           \\ \cline{2-16} 
                                      & \multicolumn{1}{c|}{\textbf{o1}} & \multicolumn{1}{c|}{\textbf{4o}} & \textbf{copilot} & \multicolumn{1}{c|}{\textbf{o1}} & \multicolumn{1}{c|}{\textbf{4o}} & \multicolumn{1}{c|}{\textbf{copilot}} & \multicolumn{1}{c|}{\textbf{o1}} & \multicolumn{1}{c|}{\textbf{4o}} & \textbf{copilot} & \multicolumn{1}{c|}{\textbf{o1}} & \multicolumn{1}{c|}{\textbf{4o}} & \multicolumn{1}{c|}{\textbf{copilot}} & \multicolumn{1}{c|}{\textbf{o1}} & \multicolumn{1}{c|}{\textbf{4o}} & \textbf{copilot} \\ \hline
\multicolumn{1}{|c|}{\textbf{Python}} & \multicolumn{1}{c|}{66\%}        & \multicolumn{1}{c|}{62\%}        & 58\%             & \multicolumn{1}{c|}{102\%}       & \multicolumn{1}{c|}{102\%}       & \multicolumn{1}{c|}{98\%}             & \multicolumn{1}{c|}{99\%}        & \multicolumn{1}{c|}{101\%}       & 97\%             & \multicolumn{1}{c|}{104\%}       & \multicolumn{1}{c|}{95\%}        & \multicolumn{1}{c|}{91\%}             & \multicolumn{1}{c|}{101\%}       & \multicolumn{1}{c|}{93\%}        & 94\%             \\ \hline
\multicolumn{1}{|c|}{\textbf{Java}}   & \multicolumn{1}{c|}{64\%}        & \multicolumn{1}{c|}{59\%}        & 51\%             & \multicolumn{1}{c|}{113\%}       & \multicolumn{1}{c|}{141\%}       & \multicolumn{1}{c|}{112\%}            & \multicolumn{1}{c|}{113\%}       & \multicolumn{1}{c|}{148\%}       & 116\%            & \multicolumn{1}{c|}{111\%}       & \multicolumn{1}{c|}{134\%}       & \multicolumn{1}{c|}{111\%}            & \multicolumn{1}{c|}{114\%}       & \multicolumn{1}{c|}{131\%}       & 112\%            \\ \hline
\multicolumn{1}{|c|}{\textbf{C++}}    & \multicolumn{1}{c|}{51\%}        & \multicolumn{1}{c|}{38\%}        & 32\%             & \multicolumn{1}{c|}{176\%}       & \multicolumn{1}{c|}{203\%}       & \multicolumn{1}{c|}{232\%}            & \multicolumn{1}{c|}{173\%}       & \multicolumn{1}{c|}{201\%}       & 234\%            & \multicolumn{1}{c|}{139\%}       & \multicolumn{1}{c|}{134\%}       & \multicolumn{1}{c|}{177\%}            & \multicolumn{1}{c|}{136\%}       & \multicolumn{1}{c|}{127\%}       & 173\%            \\ \hline
\end{tabular}
\end{table*}

\begin{figure*}[!t]
    \centering
    \includegraphics[width=\linewidth]{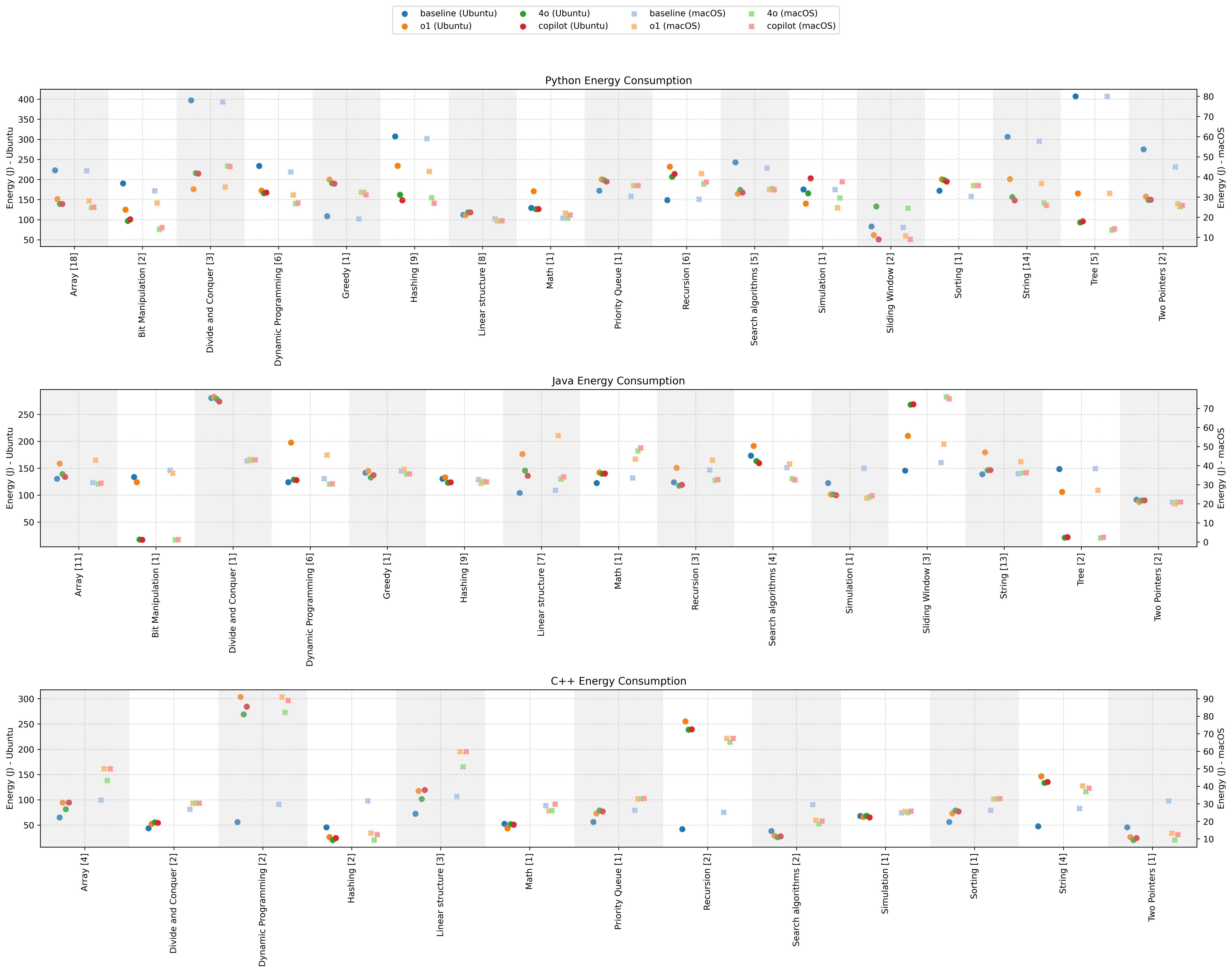}
    \caption{Each subplot illustrates the average energy consumption required to complete programming problems within each category (x-axis) for a specific programming language (Python, Java, or C++). The x-axis categories are arranged in alphabetical order. The results are displayed only for those programming problems that provided working solutions for all models. The left y-axis represents the energy consumption (in Joules) on Ubuntu, while the right y-axis represents the energy consumption (in Joules) on macOS. Scales on the y-axis are different for the three languages. The legend applies to all subplots and describes the data points for both Ubuntu and macOS.}
    \label{fig:enter-label}
\end{figure*}
\subsection{Measurement and Analysis Procedures}
For measuring energy consumption, we utilize \texttt{powermetrics} on macOS and the \textit{perf} tool on Ubuntu. On macOS, execution time, CPU and GPU power can be collected using \texttt{powermetrics}. The choice for this tool is based on its reliability, as it is one of the few available tools for macOS that provides meaningful insights into power consumption and it is developed by Apple itself and included as part of the standard macOS distribution. Since \texttt{powermetrics} reports power at regular intervals, we synchronize the execution of its process with the benchmark we want to run. For the alternative system, \textit{perf} provides access to execution time and energy consumption with \texttt{power/energy-pkg/} package. The rationale for using the \textit{perf} tool is that it offers comprehensive, generalized abstractions over hardware-specific capabilities, and is conveniently included in the linux-tools package, making it readily accessible for performance and energy measurement on Linux systems. To collect information about energy consumption, \textit{perf} leverages Intel's RAPL~\cite{rapl}, which provides reliable data about energy consumption in Intel machines~\cite{Khan:2018:RAE}.
The procedure for collecting performance and energy data followed these steps:

\begin{itemize}
    \item \textbf{Code execution}: Each code solution, whether LLM-generated or human-written, was executed under identical workload conditions across all systems. Each execution of a single solution was timed to be at least 5 seconds long, and a cool down period of 5 seconds was applied between the each execution, to ensure consistent and reliable results. 
    \item \textbf{Synchronization with Measurements}: The code solutions were executed in conjunction with the energy measurement tool to enable complete data collection. This ensures that the energy consumption data is synchronized with the execution of the code from start to finish.
    \item \textbf{Repetitions per Trial}: An execution of single was repeated 10 times to account for variability in execution and to generate robust performance data.  The exact number has been decided based on the variations of the results. Repeating the runs allowed averaging the results and mitigating any system-specific noise or anomalies. For Java, the execution was repeated 13 times, with the first 3 runs discarded to allow for JVM warm-up and ensure stable performance measurements~\cite{oopsla}. 
    \item \textbf{Data Collection and Sampling Rate}: Energy and performance data has been collected at frequent intervals of 100 samples per second (100 Hertz), ensuring fine-grained insights into the behavior of the code during execution. This sampling rate ensures that we capture both the overall energy usage and the peaks that may occur during intensive computations.
\end{itemize}
\par Following data collection, statistical and correlation analyses are conducted. To evaluate each research question, it is first necessary to determine whether the samples follow a normal distribution. So, frequency distribution plots were used to understand the type of  data distribution. We found that the data does not exhibit normality, therefore the Mann-Whitney U test was applied to test for statistical significance. Lastly, since our samples of correct code solutions were smaller than 20 in some cases, we used Hedges' $g$ to estimate the effect size. 
Spearman correlation analysis was used to examine the relationship between results in energy consumption between human and LLM-generated solutions for two machines.

\section{Results}
This section presents the results of the conducted experiments, as outlined in the methodology. All findings and conclusions are derived from benchmarks where the $\textit{p-value} <0.0001$ indicates statistical significance. This demonstrates that the energy consumption of the solutions generated by the LLMs differs statistically significantly from the baseline.
\label{sec:results}

\begin{table*}[!h]
\centering
\caption{A summary of the total number of programming problems in each group, along with the pass@1 accuracy for each programming language and model. }
\label{tab:tags_total_accuracy}
\scriptsize
\begin{tabular}{|l|c|ccc|ccc|ccc|}
\hline
\multicolumn{1}{|c|}{\multirow{2}{*}{}} & \multirow{2}{*}{\textbf{Total}} & \multicolumn{3}{c|}{\textbf{Python}}                                                            & \multicolumn{3}{c|}{\textbf{Java}}                                                              & \multicolumn{3}{c|}{\textbf{C++}}                                                               \\ \cline{3-11} 
\multicolumn{1}{|c|}{}                  &                                 & \multicolumn{1}{c|}{\textbf{o1}} & \multicolumn{1}{c|}{\textbf{4o}} & \textbf{copilot}          & \multicolumn{1}{c|}{\textbf{o1}} & \multicolumn{1}{c|}{\textbf{4o}} & \textbf{copilot}          & \multicolumn{1}{c|}{\textbf{o1}} & \multicolumn{1}{c|}{\textbf{4o}} & \textbf{copilot}          \\ \hline
\textbf{Array}                          & 38                              & \multicolumn{1}{c|}{66\%}        & \multicolumn{1}{c|}{63\%}        & 58\%                      & \multicolumn{1}{c|}{60\%}        & \multicolumn{1}{c|}{50\%}        & 47\%                      & \multicolumn{1}{c|}{45\%}        & \multicolumn{1}{c|}{37\%}        & 32\%                      \\ \hline
\textbf{Bit Manipulation}               & 13                              & \multicolumn{1}{c|}{62\%}        & \multicolumn{1}{c|}{39\%}        & 39\%                      & \multicolumn{1}{c|}{39\%}        & \multicolumn{1}{c|}{31\%}        & 39\%                      & \multicolumn{1}{c|}{23\%}        & \multicolumn{1}{c|}{15\%}        & 0\%                       \\ \hline
\textbf{Divide and Conquer}             & 3                               & \multicolumn{1}{c|}{100\%}       & \multicolumn{1}{c|}{100\%}       & 100\%                     & \multicolumn{1}{c|}{33\%}        & \multicolumn{1}{c|}{33\%}        & 33\%                      & \multicolumn{1}{c|}{100\%}       & \multicolumn{1}{c|}{67\%}        & 100\%                     \\ \hline
\textbf{Dynamic Programming}            & 17                              & \multicolumn{1}{c|}{65\%}        & \multicolumn{1}{c|}{59\%}        & 59\%                      & \multicolumn{1}{c|}{53\%}        & \multicolumn{1}{c|}{53\%}        & 47\%                      & \multicolumn{1}{c|}{47\%}        & \multicolumn{1}{c|}{29\%}        & 29\%                      \\ \hline
\textbf{Game Theory}                    & 1                               & \multicolumn{1}{c|}{100\%}       & \multicolumn{1}{c|}{0\%}         & 0\%                       & \multicolumn{1}{c|}{100\%}       & \multicolumn{1}{c|}{0\%}         & 0\%                       & \multicolumn{1}{c|}{0\%}         & \multicolumn{1}{c|}{0\%}         & 0\%                       \\ \hline
\textbf{Graph}                          & 4                               & \multicolumn{1}{c|}{50\%}        & \multicolumn{1}{c|}{0\%}         & 0\%                       & \multicolumn{1}{c|}{75\%}        & \multicolumn{1}{c|}{75\%}        & 0\%                       & \multicolumn{1}{c|}{75\%}        & \multicolumn{1}{c|}{50\%}        & 0\%                       \\ \hline
\textbf{Greedy}                         & 7                               & \multicolumn{1}{c|}{29\%}        & \multicolumn{1}{c|}{43\%}        & 43\%                      & \multicolumn{1}{c|}{57\%}        & \multicolumn{1}{c|}{57\%}        & 14\%                      & \multicolumn{1}{c|}{43\%}        & \multicolumn{1}{c|}{43\%}        & 14\%                      \\ \hline
\textbf{Hashing}                        & 16                              & \multicolumn{1}{c|}{81\%}        & \multicolumn{1}{c|}{69\%}        & 69\%                      & \multicolumn{1}{c|}{88\%}        & \multicolumn{1}{c|}{63\%}        & 63\%                      & \multicolumn{1}{c|}{63\%}        & \multicolumn{1}{c|}{45\%}        & 19\%                      \\ \hline
\textbf{Linear Structure}               & 11                              & \multicolumn{1}{l|}{72\%}        & \multicolumn{1}{l|}{90\%}        & \multicolumn{1}{l|}{90\%} & \multicolumn{1}{l|}{81\%}        & \multicolumn{1}{l|}{72\%}        & \multicolumn{1}{l|}{72\%} & \multicolumn{1}{l|}{81\%}        & \multicolumn{1}{l|}{36\%}        & \multicolumn{1}{l|}{45\%} \\ \hline
\textbf{Math}                           & 9                               & \multicolumn{1}{c|}{63\%}        & \multicolumn{1}{c|}{33\%}        & 33\%                      & \multicolumn{1}{c|}{78\%}        & \multicolumn{1}{c|}{56\%}        & 11\%                      & \multicolumn{1}{c|}{56\%}        & \multicolumn{1}{c|}{11\%}        & 11\%                      \\ \hline
\textbf{Priority Queue}                 & 3                               & \multicolumn{1}{c|}{100\%}       & \multicolumn{1}{c|}{33\%}        & 33\%                      & \multicolumn{1}{c|}{67\%}        & \multicolumn{1}{c|}{67\%}        & 0\%                       & \multicolumn{1}{c|}{100\%}       & \multicolumn{1}{c|}{100\%}       & 33\%                      \\ \hline
\textbf{Recursion}                      & 8                               & \multicolumn{1}{c|}{75\%}        & \multicolumn{1}{c|}{88\%}        & 88\%                      & \multicolumn{1}{c|}{38\%}        & \multicolumn{1}{c|}{38\%}        & 38\%                      & \multicolumn{1}{c|}{88\%}        & \multicolumn{1}{c|}{63\%}        & 38\%                      \\ \hline
\textbf{Search algorithms}              & 14                              & \multicolumn{1}{c|}{79\%}        & \multicolumn{1}{c|}{50\%}        & 50\%                      & \multicolumn{1}{c|}{86\%}        & \multicolumn{1}{c|}{57\%}        & 43\%                      & \multicolumn{1}{c|}{50\%}        & \multicolumn{1}{c|}{43\%}        & 29\%                      \\ \hline
\textbf{Simulation}                     & 1                               & \multicolumn{1}{c|}{100\%}       & \multicolumn{1}{c|}{100\%}       & 100\%                     & \multicolumn{1}{c|}{100\%}       & \multicolumn{1}{c|}{100\%}       & 100\%                     & \multicolumn{1}{c|}{100\%}       & \multicolumn{1}{c|}{0\%}         & 100\%                     \\ \hline
\textbf{Sliding Window}                 & 5                               & \multicolumn{1}{c|}{80\%}        & \multicolumn{1}{c|}{80\%}        & 40\%                      & \multicolumn{1}{c|}{80\%}        & \multicolumn{1}{c|}{60\%}        & 60\%                      & \multicolumn{1}{c|}{60\%}        & \multicolumn{1}{c|}{60\%}        & 20\%                      \\ \hline
\textbf{Sorting}                        & 7                               & \multicolumn{1}{c|}{43\%}        & \multicolumn{1}{c|}{29\%}        & 29\%                      & \multicolumn{1}{c|}{43\%}        & \multicolumn{1}{c|}{29\%}        & 0\%                       & \multicolumn{1}{c|}{43\%}        & \multicolumn{1}{c|}{43\%}        & 29\%                      \\ \hline
\textbf{String}                         & 23                              & \multicolumn{1}{c|}{87\%}        & \multicolumn{1}{c|}{78\%}        & 78\%                      & \multicolumn{1}{c|}{70\%}        & \multicolumn{1}{c|}{65\%}        & 61\%                      & \multicolumn{1}{c|}{57\%}        & \multicolumn{1}{c|}{35\%}        & 48\%                      \\ \hline
\textbf{Tree}                           & 9                               & \multicolumn{1}{c|}{78\%}        & \multicolumn{1}{c|}{67\%}        & 78\%                      & \multicolumn{1}{c|}{67\%}        & \multicolumn{1}{c|}{56\%}        & 56\%                      & \multicolumn{1}{c|}{56\%}        & \multicolumn{1}{c|}{22\%}        & 44\%                      \\ \hline
\textbf{Two Pointers}                   & 2                               & \multicolumn{1}{c|}{100\%}       & \multicolumn{1}{c|}{100\%}       & 100\%                     & \multicolumn{1}{c|}{100\%}       & \multicolumn{1}{c|}{100\%}       & 100\%                     & \multicolumn{1}{c|}{50\%}        & \multicolumn{1}{c|}{50\%}        & 50\%                      \\ \hline
\end{tabular}
\end{table*}
\subsection{Variations in the energy-efficiency of the LLM-generated code between Python, Java and C++.}
\par Although \textbf{RQ1} explicitly focuses on the aspect of energy efficiency, this study also considers pass@1 accuracy and execution time as complementary metrics. Table \ref{tab:accuracy_energy_exec} summarizes the data for each programming language, model, and machine across these metrics. The arrows accompanying each metric denote whether a higher or lower value is considered preferable. The pass@1 accuracy metric has an upper bound of 100\%. Conversely, the percentages reported for energy consumption and execution time represent the proportion of energy and time utilized relative to the baseline. Thus, a lower value signifies that the generated solutions are, on average, more energy-efficient or faster relative to the baseline, whereas a higher value indicates higher energy consumption or increased execution time. 
\par Table \ref{tab:accuracy_energy_exec} demonstrates a consistent trend in pass@1 accuracy, with Python achieving the highest values, followed by Java and C++ across all models. The highest recorded value is 66\% for Python generated by the OpenAI o1-mini model, while the lowest is 32\% for C++ produced by GitHub Copilot. In terms of energy efficiency, Python solutions generated by LLMs were, in certain cases, more energy-efficient than the baseline. The most noticeable reductions were observed on macOS systems for both GitHub Copilot and GPT-4o, with GitHub Copilot additionally exhibiting reduced energy consumption on Ubuntu system. However, in the remaining cases for Python, the differences compared to the baseline were negligible, with the largest observed variation being a 4\% increase on macOS using OpenAI o1-mini. Execution time for Python solutions also exhibited a decrease relative to the baseline, particularly on macOS with GitHub Copilot and GPT-4o, as well as on the Ubuntu system with GitHub Copilot. Overall, the relationship between energy consumption and execution time remains linear across all programming languages, systems, and machines.

\par On the other hand, an increase in energy consumption is observed for Java and C++ across all models and systems. Although an increase is observed for two languages, the increase in Java is less pronounced compared to C++. The most significant increase in Java was recorded using GPT-4o, with rises of 41\% on the Ubuntu system and 34\% on the macOS system. The results for OpenAI o1-mini and GitHub Copilot are not significantly different from the baseline on either platform. The increase for C++ is notably more pronounced, with the highest recorded rise of 132\% observed when using GitHub Copilot on the Ubuntu system. Additionally, other models also generated solutions with higher energy consumption compared to the baseline, with the smallest increase of 34\% noted on macOS with GPT-4o. Since the relationship between execution time and energy consumption remains linear, the execution time for both languages is also consistently longer compared to the baseline.
\begin{summary}
{\footnotesize 
\textbf{Summary.} 
The results indicate that the models exhibit
superior performance in code generation tasks for Python, achieving the highest pass@1 accuracy among the three languages, while results for C++ demonstrate the lowest accuracy. With respect to energy consumption, the models generated solutions comparable to the baseline for Python and Java, and in certain instances, Python solutions proved to be more energy-efficient and faster than human-written counterparts. However, the solutions generated for C++ are significantly more energy-intensive compared to the baseline.
}
\end{summary}

\begin{figure*}
    \centering
    \includegraphics[width=\linewidth]{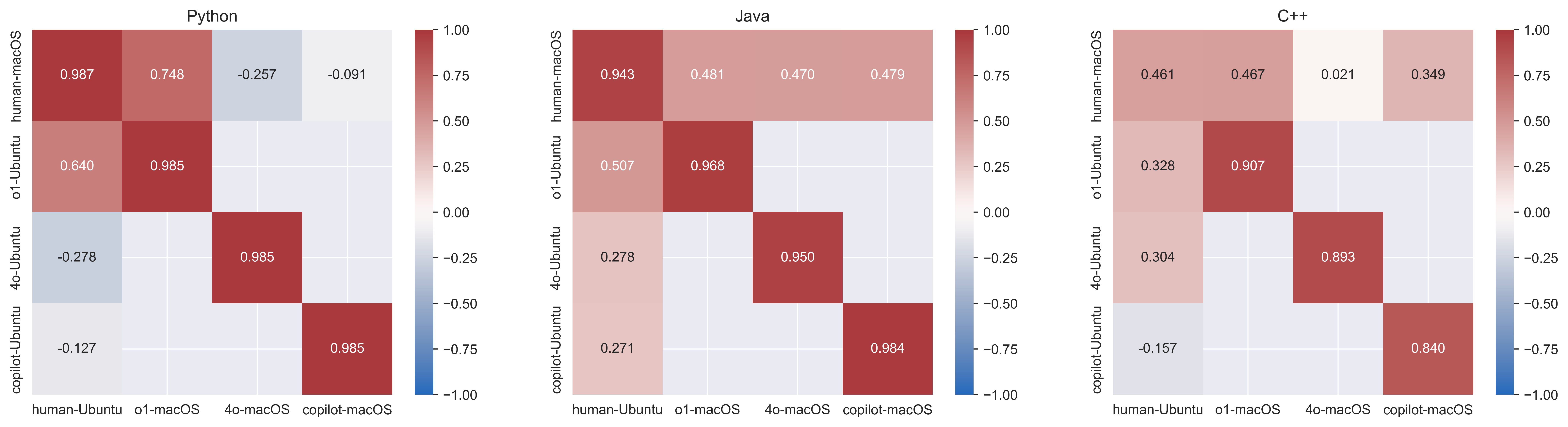}
    \caption{Illustration of the Spearman correlation between the energy consumption results for solutions generated by each model across two platforms. }
    \label{fig:correlation}
\end{figure*}
\subsection{Impact of data structure and algorithmic technique selection on the energy-efficiency of the LLM-generated code.}
Figure \ref{fig:enter-label} illustrates  the average energy consumption required to complete programming problems within each category for each programming language in this study. It only includes the categories where all models successfully generated solutions that passed the predefined tests. A comprehensive list of all data structures and algorithmic techniques, along with their corresponding pass@1 scores, is provided in Table \ref{tab:tags_total_accuracy}.
\par Figure \ref{fig:enter-label} reveals that LLM-generated solutions for Python exhibit higher energy efficiency compared to the baseline for specific data structures, including \textit{Array}, \textit{String}, and \textit{Tree}. For other data structures, the results generally align closely with the baseline, though more notable deviations are observed for \textit{Priority Queue}. Regarding algorithms, the Python solutions demonstrate significantly higher efficiency for \textit{Bit Manipulation}, \textit{Dynamic Programming}, \textit{Divide and Conquer}, \textit{Search algorithms}, \textit{Two Pointers}, and \textit{Hashing}. For other algorithmic techniques, the results are largely consistent with the baseline, with more pronounced deviations that are more energy consuming for \textit{Recursion}, \textit{Greedy} and \textit{Sorting}. 
\par For Java, the results concerning data structures are largely comparable to the baseline, with a notably higher energy consumption observed for \textit{Linear structures} and greater energy efficiency for \textit{Tree}-based structures. Among the algorithms, \textit{Math} and \textit{Sliding Window} show significantly increased energy usage, whereas \textit{Bit Manipulation} and \textit{Simulation} demonstrate improved efficiency. Other algorithmic techniques and data structures display performance that is generally comparable to the baseline with some differences between the model performances. It is evident that OpenAI o1-mini generates solutions with higher energy consumption for \textit{Dynamic Programming}, \textit{Recursion}, \textit{Search Algorithms}, \textit{String}, and \textit{Array} tasks. In contrast, GitHub Copilot and GPT-4o produce more energy-efficient solutions for these categories compared to the baseline. However, the observed differences are not significant.
\par Finally, for C++, noticeably fewer programs are presented in the groups, reflecting the models' more limited code generation abilities for this language. Overall, the energy consumption of the generated solutions is generally higher compared to the baseline, with the exception of a few algorithms, including \textit{Hashing}, \textit{Search Algorithms}, and \textit{Two Pointers}, demonstrating improved efficiency. The most significant increases in energy consumption are observed for \textit{Dynamic Programming}, \textit{Linear Structure}, \textit{Recursion}, \textit{Sorting}, and \textit{String}. 
\par  To summarize the insights across programming languages, the performance of specific data structures and algorithmic techniques varies by language. For instance, as shown in Table \ref{tab:tags_total_accuracy}, the pass@1 accuracy for \textit{String}, \textit{Linear Structure} and \textit{Tree}, remains consistently high across all languages. As evident from the previously discussed results, the energy consumption for \textit{String} and \textit{Tree} tasks remains either below the baseline or closely aligned with it. However, the same observation does not hold for \textit{Linear Structures}, where energy consumption exhibits a notable increase for Java and C++, while remaining consistent with the baseline for Python. This indicates that high correctness in solution generation does not necessarily correlate with energy efficiency. Another notable observation pertains to \textit{Hashing} and \textit{Search algorithms}. Both of these categories exhibit consistently high pass@1 accuracy across all languages and models. Furthermore, the previously discussed findings indicate that LLM-generated solutions exhibited either improved efficiency or performance comparable to the baseline across both categories.

\par Regarding the limitations of LLM-generated solutions, programming problems associated with \textit{Sorting} consistently exhibit challenges for the models, showing low pass@1 accuracy across all programming languages and models. Similar difficulties are observed for problems involving \textit{Graph} structures and \textit{Greedy algorithmic techniques}. For the latter, energy consumption is also notably higher compared to human-written solutions. 
Moreover, programming problems associated with \textit{Math} also posed challenges for code generation, exhibiting higher energy consumption across all programming languages. This trend is observed even for Python, which otherwise consistently demonstrates higher energy efficiency compared to the baseline across the majority of algorithmic techniques and data structures. Lastly, \textit{Recursion} can also be categorized among the tasks that exhibit higher energy consumption during code generation.
\begin{summary}
{\footnotesize 
\textbf{Summary:} \textit{String} and \textit{Tree} exhibit consistently high pass@1 accuracy across all languages, reflecting their suitability for LLM-generated code. \textit{Hashing} operations and \textit{Search algorithms} also demonstrate high pass@1 accuracy and enhanced energy efficiency in some cases. However, challenges remain for \textit{Sorting}, \textit{Graph}, and \textit{Greedy algorithmic techniques}, which exhibit low pass@1 accuracy and higher energy consumption, especially for the latter. \textit{Math} problems and \textit{Recursion} also pose difficulties, with higher energy consumption observed, particularly for Java and even Python in some cases.
}
\end{summary}

\subsection{Effects of the chosen LLM on the energy footprint of the generated code.}
\begin{figure*}
    \begin{minipage}{0.48\textwidth}
    \begin{tcolorbox}[title= OpenAI o1-mini solution, colframe=myblue]
    \begin{lstlisting}[style=mystyle]
def firstMissingPositive(nums):
    n = len(nums)
    for i in range(n):
        while 1 <= nums[i] <= n 
            and nums[nums[i] - 1] != nums[i]:
             nums[nums[i] - 1], nums[i] = nums[i], nums[nums[i] - 1]
    for i in range(n):
        if nums[i] != i + 1:
            return i + 1
    return n + 1
    \end{lstlisting}
    \end{tcolorbox}
\end{minipage}
\hfill
\begin{minipage}{0.48\textwidth}
    \begin{tcolorbox}[title=Best human-written solution, colframe=myblue]
    \begin{lstlisting}[style=mystyle]
def firstMissingPositive(nums):
    nums.append(0)
    n = len(nums)
    for i in range(len(nums)):
        if nums[i]<0 or nums[i]>=n:
            nums[i]=0
    for i in range(len(nums)):
        nums[nums[i]%n] += n
    for i in range(1, len(nums)):
        if nums[i]/n == 0: return i
    return n
    \end{lstlisting}
    \end{tcolorbox}
\end{minipage}
\caption{Example solutions for the \textit{First Missing Positive} problem. The left solution was generated by OpenAI's o1-mini model, while the right solution represents the highest-rated human implementation from LeetCode. }
\label{fig:examplePythonCode}
\end{figure*}
\par Table \ref{tab:accuracy_energy_exec} reveals the first key insight: the accuracy of code generation has generally improved in OpenAI o1-mini compared to GPT-4o and GitHub Copilot across all languages. Additionally, it is noteworthy that GitHub Copilot's performance lags behind GPT-4o, highlighting the differences between the two models, despite both being based on the GPT-4 series. For a more detailed analysis, Table \ref{tab:tags_total_accuracy} highlights that OpenAI o1-mini has shown significant improvements in categories such as \textit{Search Algortihms}, \textit{Bit Manipulation}, \textit{Math}, \textit{Game Theory}, \textit{Graph}, and \textit{Sorting}. While improvements are observed across all categories, these stand out as particularly notable because other models frequently failed to generate working solutions, with pass@1 scores often recorded at 0. These improvements underscore the incorporation of reinforcement learning in OpenAI o1-mini, enabling the model to engage in reasoning through its chain-of-thought mechanism.
\par In terms of energy efficiency, a different trend emerges: solutions generated by OpenAI o1-mini consume, on average, more energy than those produced by GPT-4o and GitHub Copilot. The latter out of the two is the most efficient model for Python and Java, while GPT-4o demonstrates better efficiency for C++. This trend is consistent when examining execution time. Furthermore, the average energy consumption of GPT-4o and GitHub Copilot is similar, emphasizing the comparable performance of the two models.
\par The difference between the newer OpenAI o1-mini version and the older GPT-4o version can also be observed by Figure \ref{fig:enter-label}. It only encapsulates the programming tasks that were produced correctly by all the models, providing a fair comparison in terms of energy consumption between the three. In many categories, OpenAI o1-mini typically consumes more energy than both GPT-4o and GitHub Copilot across all languages. However, there are instances where OpenAI o1-mini exhibits slightly better energy efficiency, such as in the \textit{Divide and Conquer} and \textit{Sliding Window}. Nonetheless, in general the energy consumption of solutions generated by OpenAI o1-mini exhibits higher values compared to its predecessor models.
\par Another significant observation is that in certain cases, the energy efficiency of solutions generated by OpenAI o1-mini closely approximates that of human-written solutions. To evaluate this observation, we refer to Figure \ref{fig:correlation}, which presents the Spearman correlation coefficients for energy consumption among the solutions generated by the models. The analysis reveals that the correlation coefficients between the energy consumption of human solutions and those generated by OpenAI o1-mini are higher for each language compared to the coefficients between human solutions and those generated by GPT-4o or GitHub Copilot. For Python, the correlation ranges from 0.748 to 0.640, depending on the platform, indicating a significant relationship between the results. However, for Java and C++, the correlation reduces, occasionally aligning with the values observed for the other models.
\par To provide a comparison between the solution generated by OpenAI's o1-mini model and the best human-written solution, Figure \ref{fig:examplePythonCode} presents both approaches for the \textit{First Missing Positive} problem on LeetCode. This problem requires identifying the smallest positive integer missing from a given unsorted integer array \texttt{nums}. The human-written solution achieves an $O(n)$ time complexity, whereas the OpenAI o1-mini-generated solution initially appears to exhibit an $O(n^2)$ complexity. However, it employs an in-place swapping technique, which efficiently positions the elements without unnecessary operations. Furthermore, the number of iterations within the nested loop of the OpenAI o1-mini-generated solution is relatively low, traversing the list approximately 1.5 times, with swaps occurring only sporadically rather than in every iteration. In contrast, the human-written solution relies on additional computations, such as modulus and division operations, which are executed during every iteration. This solution can process the list roughly 2.5 times, potentially leading to higher computational overhead.
\begin{summary}
{\footnotesize 
\textbf{Summary:}  While OpenAI o1-mini demonstrates notable improvements in code generation pass@1 accuracy across languages, with notable improvements in categories such as \textit{Search Algortihms}, \textit{Bit Manipulation}, \textit{Math}, \textit{Game Theory}, \textit{Graph}, and \textit{Sorting}, it generally consumes more energy than GPT-4o and GitHub Copilot. GitHub Copilot generates the most energy-efficient solutions for Python and Java, while GPT-4o excels in C++. Despite their higher energy consumption, OpenAi o1-mini's solutions in some cases approach the energy efficiency of human-written solutions, as evidenced by higher correlation coefficients for Python. However, this correlation weakens for Java and C++, but still remains higher than for GPT-4o and GitHub Copilot.
}
\end{summary}

\subsection{Variations in energy efficiency of LLM-generated code on different platforms.}
This study examines the energy consumption of LLM-generated code on two platforms, an Apple M3 Mac running macOS Sonoma and a Lenovo PC running Ubuntu 22.04. These systems are referred to as macOS and Ubuntu throughout the text. The selection of these platforms was guided by the availability of reliable energy measurement tools, namely \texttt{powermetrics} for macOS and \texttt{perf} for Ubuntu. 
\par Figure \ref{fig:correlation} illustrates the correlation between energy consumption results obtained on the two platforms. The data reveals a strong correlation for the baseline results in Python and Java, whereas C++ exhibits a lower correlation coefficient of 0.461. It is important to note that each solution for C++ was compiled independently on each machine, which can be the reason for the difference in the energy footprint~\cite{compiler_opt}. 
\par Although the baseline results show a low correlation score across the two platforms, this is not the case for LLM-generated solutions. Notably, the correlation between LLM-generated solutions executed on Ubuntu and the same solutions recompiled and executed on macOS is high, with values ranging between 0.9 and 0.8, depending on the model. Figure \ref{fig:enter-label} also illustrates that the efficiency patterns of LLM-generated solutions remain consistent across systems, with solutions being either more or less efficient in the same categories regardless of the platform. Based on results, we can hypothesize that LLM-generated solutions may be machine-agnostic, whereas human-written solutions from LeetCode could be optimized for the user's specific machine, with a greater proportion of users potentially favoring either Ubuntu or macOS. 
\par Regarding energy consumption, Table \ref{tab:accuracy_energy_exec} illustrates a trend where the average increase in energy consumption relative to the baseline is more pronounced on the Ubuntu platform across all languages. However, the difference between the two systems is not significant for Python and Java, whereas for C++, the discrepancy is more visually apparent, though still not statistically significant. Overall, Figure \ref{fig:enter-label} demonstrates that the patterns of energy efficiency in LLM-generated solutions, whether more or less efficient, remain consistent across the two machines. 
\begin{summary}
{\footnotesize 
\textbf{Summary:}  As observed, none of the models exhibit a preference for one machine over the other, displaying nearly identical energy consumption patterns across both systems. LLM-generated solutions exhibit a strong correlation (0.9-0.8) across both systems, indicating that they are machine-agnostic, whereas human-written solutions may appear to be optimized for specific machines, as evidenced by the low correlation between human solutions for C++. 
}
\end{summary}

\section{Threats to Validity}
\label{sec:threats}
\textbf{Construct Validity.} In this study, our research question focuses on analyzing the energy efficiency of LLM-generated code. Relying on a single metric to address the question of energy efficiency could jeopardize the validity of the findings. Therefore, rather than basing conclusions solely on power consumption and energy usage, we also tracked additional metrics such as execution time and correctness, as these factors are also critical to the overall efficiency of the code. \\ Another potential threat to validity arises from the outdated nature of some human-written solutions, many of which were posted nearly a decade ago. To mitigate this, we selected solutions with a newer version of the language that preserves the implementation approach of the original solution. Those solutions are usually posted in the same thread in the comments section. \\ Lastly, to reduce the risk of interaction between the different treatments, we allowed the system rest for 30 seconds between measurements to allow it to cool down. 

\textbf{Internal Validity.} 
For instance in our study, the variability in system performance, background processes, or system settings (e.g., power-saving modes) had the potential to distort the results. Additionally, the accuracy of the tools used to measure energy consumption and performance posed a risk, as measurement overhead could have affected the results. We minimized this threat by controlling all relevant system settings, including disabling power-saving modes, ensuring consistent battery levels, and reducing background processes. Additionally, we ran the tests multiple times to account for the variability and to average out potential anomalies. 

\textbf{External Validity.}  The LLMs in our study were trained on publicly available internet data, possibly including LeetCode. Studies show LLMs perform well on "easy" problems but struggle with harder ones, so we selected "hard" problems. To ensure generalizability across programming languages, we chose three widely used languages that highlight potential inefficiencies in LLM-generated code.

\textbf{Conclusion Validity.} A potential threat to conclusion validity is the possibility of drawing incorrect conclusions due to small sample sizes, variability in the data, or inappropriate statistical tests. In our study, constructing a sufficiently large benchmark was essential for drawing reliable conclusions, especially about the differences in energy efficiency regarding programming approaches. This posed a challenge due to the limited availability of human-written solutions for "hard" problems. Since the sample size proved to be on the smaller side, we employed statistical methods specifically designed for small sample sizes (e.g., Mann-Whitney U test) to ensure the validity of the analysis. We further mitigated this threat by running multiple trials of each experiment to ensure sufficient data for robust statistical analysis.

\section{Conclusion}
\par The primary objective of this research is to evaluate the current capabilities of GitHub Copilot, GPT-4o, and OpenAI o1-mini in generating energy-efficient code across three programming languages, executed on two distinct operating systems: Ubuntu 22.04 and macOS Sonoma v14.6. The study highlights that LLMs perform best in Python, achieving the highest pass@1 accuracy. With respect to energy efficiency, the models yield results comparable to the baseline for Python and Java, with Python solutions, in some cases, demonstrating greater energy efficiency.
\textit{String} and \textit{Tree} related tasks along with \textit{Hashing} operations and \textit{Search algorithms} consistently excel, but challenges remain in \textit{Sorting}, \textit{Graph}, \textit{Greedy algorithmic techniques}, and tasks involving \textit{Math}, and \textit{Recursion}. The OpenAI o1-mini model shows significant accuracy improvements, particularly in \textit{Search algorithms} and \textit{Sorting}, but consumes more energy than GPT-4o and GitHub Copilot across all aspects. Additionally, our findings indicate that the energy efficiency of OpenAI o1-mini-generated solutions closely aligns with that of human-written code. In regard to the platforms, LLM-generated solutions are machine-agnostic, showing strong energy correlation across systems, unlike human solutions, which appear optimized for specific machines, as seen in low correlation for C++.
\par \textbf{Implications.} 
While the findings indicate improved energy efficiency in LLM-generated solutions for Python, it is important to note that Python is a language where application performance and energy consumption are generally less critical in practical scenarios. On the other hand, the results for Java and C++ are less promising, suggesting that, in practice, it would be prudent to carefully evaluate LLM-generated code for efficiency to ensure that energy consumption remains within acceptable limits.
\label{sec:conclusion}

\bibliographystyle{IEEEtran}
\bibliography{references}
\end{document}